\documentclass[pra,reprint,showpacs,twocolumn,superscriptaddress,floatfix]{revtex4-1}

\usepackage{epsfig}
\usepackage{amsmath}
\usepackage{amssymb}
\usepackage{amsfonts}
\usepackage{mathptmx}
\usepackage{dcolumn}
\usepackage{eucal}
\usepackage{bm}
\usepackage{color}
\usepackage[colorlinks,linkcolor=blue,citecolor=blue]{hyperref}
\usepackage{epstopdf}
\usepackage{bbold}
\usepackage{url}
\usepackage{mathtools}
\usepackage{dsfont}

\usepackage{dsfont}

\newcommand{\average}[1]{\langle #1 \rangle}
\newcommand{\Trace}{{\rm Tr}}

\def \ket#1{\mathinner{|{#1}\rangle}}
\def \bra#1{\mathinner{\langle{#1}|}}

\def\braket#1{\mathinner{\langle{#1}\rangle}}

\newcommand{\ketbra}[2]{{\mathinner{| {#1} \rangle \langle {#2} |}} }

\newcommand{\matrixel}[3]{{\mathinner{\langle{#1}| {#2} | {#3}\rangle}} }

\newcommand{\GFunc}{\mathcal{G}_\lambda}

\newcommand{\mU}{\mathcal{U}}

\newcommand{\ej}[1]{{\bf e}_{j_#1}}
 
\begin{document}

\title{Quantum gradient evaluation through quantum non-demolition measurements}

\author{P. Solinas}
\affiliation{Dipartimento di Fisica, Universit\`a di Genova, via Dodecaneso 33, I-16146, Genova, Italy}
\affiliation{INFN - Sezione di Genova, via Dodecaneso 33, I-16146, Genova, Italy}
%\email{}
\author{S. Caletti}
\affiliation{Dipartimento di Fisica, Universit\`a di Genova, via Dodecaneso 33, I-16146, Genova, Italy}
\affiliation{INFN - Sezione di Genova, via Dodecaneso 33, I-16146, Genova, Italy}
\author{G. Minuto}
\affiliation{Dept. of Informatics, Bioengineering, Robotics, and Systems Engineering (DIBRIS), Polytechnic School of Genoa University, Genova, Italy}
\affiliation{INFN - Sezione di Genova, via Dodecaneso 33, I-16146, Genova, Italy}

\date{\today}

%\pacs{85.80.Lp,74.50.+r,72.25.-b}
\begin{abstract}
We discuss a Quantum Non-Demolition Measurement (QNDM) protocol to estimate the derivatives of a cost function with a quantum computer. %This is a key step for the implementation of variational quantum circuits.
The cost function, which is supposed to be classically hard to evaluate, is associated with the average value of a quantum operator. Then a quantum computer is used to efficiently extract information about the function and its derivative by evolving the system with a so-called variational quantum circuit.
To this aim, we propose to use a quantum detector that allows us to directly estimate the derivatives of an observable, i.e., the derivative of the cost function.
With respect to the standard direct measurement approach, this leads to a reduction of the number of circuit iterations needed to run the variational quantum circuits.
The advantage increases if we want to estimate the higher-order derivatives.
We also show that the presented approach can lead to a further advantage in terms of the number of total logical gates needed to run the variational quantum circuits.
These results make the QNDM a valuable alternative to implementing the variational quantum circuits.
\end{abstract}

\maketitle
 
\section{Introduction}\label{sec1}

The advent of all-purpose quantum computers able to solve hard computational problems is still decades away.
However, it is commonly believed that some problems intractable with a classical computer could be within reach of today's Noisy Intermediate-Scale Quantum (NISQ) computers. 

The problems for which a quantum advantage can be reached in the next years are the ones that require a large working space such as the simulation of complex physical and chemical systems.
The most promising architecture is a hybrid quantum-classical one.
Among these hybrid algorithms the most relevant ones are the variational quantum circuit, the variational quantum eigensolver \cite{Peruzzo2014, Kandala2017} and the quantum approximate optimization algorithm \cite{farhi2014}.
The computational scheme is the following.
Given a certain cost function in a large parameter space to be minimized, we associate it with a Hermitian operator.
We run a quantum circuit to evaluate the average values of the Hermitian operator, i.e., the value of the cost function, in a specified point of the parameter space.
Since the quantum measurements are probabilistic, we need to iterate the process to reach the desired accuracy.
This information is then fed into a classical computer which elaborates it and determines the following steps of the quantum computer.
Typically, classical computation consists in an optimization algorithm for which we need information about the derivatives of the cost functions.
Therefore, in these schemes, the main quantum computational task is to obtain the derivatives of the cost function by measuring quantum observables with accuracy and minimal cost. 
From the quantum perspective, the main resource costs come from the iterations needed to have an accurate average and the logical gates needed to run the quantum circuit at every iteration.
Because of the limitation in quantum hardware and quantum operations, any reduction of the cost can bring us closer to obtaining a quantum advantage in problems of practical interest.
 
While the usual proposals rely on the direct measurement (DM) of a quantum observable to extract information about the cost function \cite{McArdle2020,Cerezo2021,Mari2021}, here, we discuss an alternative method to estimate directly the derivatives of the observable.
A quantum detector is coupled in sequence with the system from which we want to extract the information.
The information about the observable, its gradient, or the high derivatives is stored in the detector phase which is eventually measured.
This technique is often called Quantum Non-Demolition Measurement (QNDM) \cite{clerk2011full} or full counting statistics  \cite{bednorz2012nonclassical} and it is rooted in the idea of weak values and weak measurements \cite{Aharonov1988weak}.

The potential advantages of this approach lie in the fact that, with this unconventional measurement, we can directly estimate the average of the {\it variation}, i.e., the gradient, of quantum observables which cannot be obtained with direct measurement.
Indeed, the same approach has been used to estimate the variation of charge \cite{bednorz2012nonclassical} and energy \cite{solinas2015fulldistribution,solinas2016probing,solinas2021,solinas2022} in quantum systems.
Since these are related to observables at different times, they are not associated with any hermitian operator  \cite{talkner2007fluctuation} and the measurement process suffers from conceptual and practical subtleties \cite{solinas2015fulldistribution,solinas2022}.
In this paper, we revert the process and identify in the QNDM the ideal approach to extract information about the derivative of an observable.

A similar approach was presented in Refs. \cite{OBrien2019, Banchi2021}.
Here, we discuss it in a more general framework, extend its applications to estimate high derivatives and discuss the main advantages with respect to the direct measurement approach.
As we show below, the QNDM needs fewer resources (in terms of iterations and logical operators)  than the DM approach. The advantages increase with the order of the derivatives to be estimated.

The paper is divided as follows.
In Sec.~\ref{sec:intro}, we introduce the quantities to be estimated.
In Sec.~\ref{sec:gradient}, we show how the gradient of an observable can be measured with the QNDM approach and, in Sec.~\ref{sec:example_and_resources} we give an explicit example of the advantages of the QNDM approach when the cost function is related to a complex operator. In Sec.~\ref{sec:higher_derivatives}, we extend our framework to the estimate of second and higher derivative and in Sec.~\ref{sec:advantages} we discuss the advantages in terms of resources needed. Finally, Sec.~\ref{sec:conclusions} contains the conclusions.

\section{Definition of the fundamental quantities}
\label{sec:intro}

Let's suppose to have a quantum state $\lvert\psi_0\rangle$ and we let it evolve with a unitary operator $U(\theta)$ which depends on $m$ parameters ${\vec \theta} = ( \theta_1, \theta_2, .... , \theta_m )$. The final state is $\lvert \psi ({\vec \theta})\rangle = U({\vec \theta}) \lvert\psi_0\rangle$.
In a more physical (and less abstract) framework, we can consider the operator $U({\vec \theta})$ generated by the hamiltonian $H({\vec \theta})$  depending on the $m$ parameters which can depend on time ${\vec \theta}(t) = (\theta_1(t), \theta_2(t), ... , \theta_m(t) )$.
In this case, the evolution is given by $U({\vec \theta}) = T \exp\{ -i \int_0^\tau H({\vec \theta}(t)) d t\}$ where $T$ is the time-ordered product.

To make a specific example for quantum computers, we can consider the system composed of $n$ qubits.
Notice that, in general, we have $n \neq m$ so that we have enough degrees of freedom to implement different unitary transformation  $U({\vec \theta})$.
The initial state is $\lvert\psi_0\rangle= \lvert0\rangle = \lvert00....0\rangle$, and the unitary transformation $U(\theta)$ can be obtained as a sequence of quantum logical gates.

We assume that the $U({\vec \theta})$ operator can be written as 
\begin{equation}
 	U({\vec \theta}) = V_m U_m({\vec \theta_m})....V_2 U_2({\vec \theta_2}) V_1 U_1({\vec \theta_1})
	\label{eq:U_def}
\end{equation}
where $V_m$ are arbitrary transformation (independent of $\vec \theta$) and $U_m({\vec \theta_j}) = \exp{\{- i \theta_j H_j \}}$ is a single qubit operator generated by the Hermitian operator $H_j$ \cite{Mari2021}.
Supposing that  $H_j^2 =   \mathds{1}$,
\begin{equation}
	U_j({\vec \theta_j}) = e^{- i H_j \theta_j/2} = \cos \frac{\theta_j}{2} \mathds{1} -i \sin \frac{\theta_j}{2} H_j.
\end{equation}
A more general situation where the $H_j$s do not satisfy the above condition is discussed in Ref. \cite{Banchi2021}.

The quantity we are interested in is the average value of another operator $\hat{M}$ over the final state $\lvert\psi ({\vec \theta})\rangle$ \cite{Mari2021}
\begin{equation}
	f({\vec \theta}) = \langle\psi ({\vec \theta})\lvert\hat{M}\vert\psi ({\vec \theta})\rangle = \langle0\lvert U^\dagger({\vec \theta}) \hat{M} U({\vec \theta})\lvert0\rangle.
	\label{eq:f_def}
\end{equation}

Measuring this quantity for two points, we can estimate the derivative of $f({\vec \theta})$ as \cite{Mari2021}
\begin{equation}
	g_{j_1} = \frac{\partial f({\vec \theta})}{\partial \theta_{j_1}} = \frac{f(\vec \theta +s \ej{1} ) - f(\vec \theta -s \ej{1})}{ 2 \sin s }.
	\label{eq:f_grad_def}
\end{equation}
where $\ej{1}$ is the versor along the $\theta_{j_1}$ direction.
This method can be generalized to calculate the second derivative (see Eq. ($10$) in  \cite{Mari2021})
\begin{equation}
\begin{split}
    g_{j_1,j_2} &= \frac{\partial^2 f({\vec \theta})}{\partial \theta_{j1} \partial \theta_{j2}} = \\ 
    &=\Big[ f(\vec \theta +s (\ej{1}+\ej{2})) - f(\vec \theta +s (-\ej{1}+\ej{2})) \\
    &-f(\vec \theta +s (\ej{1}-\ej{2})) + f(\vec \theta -s (\ej{1}+\ej{2}))\Big] \\   
    &\Big[2 \sin^2 s \Big]^{-1}
\label{eq:second_derivative}
\end{split}
\end{equation}
or even higher derivatives defined as \cite{Mari2021} 
\begin{equation}
	g_{j_1, j_2,...,j_d} = \frac{\partial^d f(\vec \theta)}{\partial \theta_{j1} \partial \theta_{j_2}...\partial \theta_{j_d}}.
\end{equation}

The information about the derivatives of $f$ can be used to minimize a cost function.
To this aim, different optimizer algorithms can be used (see, for example, \cite{Mari2021}).
The most common one is the gradient descent (GD). 
This optimiser starts with the parameter $\vec \theta^{(0)}$ which are sequentially updated to new values $\vec \theta^{(1)}$, $\vec \theta^{(2)}$, ..., $\vec \theta^{(T)}$.
The update of parameters is obtained by the rule $\vec \theta^{(t)} = \vec \theta^{(t-1)} - \eta \nabla f (\vec \theta^{(t-1)})$ where $\eta >0 $.

The access to the second derivatives allows us to implement other optimization algorithms.
For example, in the Newton optimizer, the parameter update rule is $\vec \theta^{(t)} = \vec \theta^{(t-1)} - \eta [H f (\vec \theta^{(t-1)})]^{-1} \nabla f (\vec \theta^{(t-1)})$ where $[H f (\vec \theta^{(t-1)})]^{-1}$ is the inverse of the Hessian matrix estimated using Eq.~(\ref{eq:second_derivative}).

Since the Newton optimizer requires the inversion of the Hessian matrix, it might be resource-consuming for a large parameter space.
For this reason, alternative second-order approaches have been proposed such as the diagonal Newton optimizer and quantum natural gradient optimizer \cite{cheng2010, Liu2019, Mari2021}.
The latter uses the information encoded in the Fubini-Study metric tensor that can be obtained with similar methods \cite{Mari2021}.

\section{Quantum non-demolition measurement: First order derivative - Gradient}
\label{sec:gradient}

The key idea of the QNDM approach is to couple the system to a quantum detector and store the desired information in the phase of the detector that is eventually measured \cite{OBrien2019, Banchi2021}.
By choosing properly the system detector interaction \cite{solinas2015fulldistribution,solinas2016probing,SolinasPRA2017}, we can store in the detector the information of the {\it variation of the average $M$} so that a final measurement gives us direct access to the gradient~(\ref{eq:f_grad_def}).
 
 Notice that in terms of operators we can write Eq.~(\ref{eq:f_grad_def}) as the average of a new operator $
 U^\dagger (\vec \theta +s \ej{1} ) \hat{M} U(\vec \theta +s \ej{1} ) - U^\dagger (\vec \theta -s \ej{1} ) \hat{M} U(\vec \theta -s \ej{1} )
 = \hat{M}(\vec \theta +s \ej{1} ) - \hat{M} (\vec \theta -s \ej{1} )$.
However, being this the difference between observables, it is not a quantum observable and it cannot be measured directly but the contributions must be measured separately \cite{talkner2007fluctuation, solinas2015fulldistribution,solinas2016probing}.
The QNDM approach allows us to access the same information about the averages with direct measurement of the detector phase, therefore, reducing the resources needed.

More precisely, let us consider the same setup discussed in Sec.~\ref{sec:intro} but let us add an additional quantum system that acts as the detector.
We suppose that the detector has no dynamics, i.e., it effectively evolves over times much longer than the one needed to perform the full protocol.

The system and the detector interact with the Hamiltonian
\begin{equation}
	H_{I} = h(t) \lambda \hat{p} \otimes \hat{M}.
\end{equation}
Here, $\lambda$ is the coupling constant between the system and the detector, and it can be changed and fixed at the beginning of the evolution. The operators $\hat{p}$ and $\hat{M}$ act on the detector and the system, respectively, and we denote the eigenstates of $\hat{p}$ with $\lvert{p}\rangle$, i.e., $\hat{p} \lvert{p}\rangle = p \vert{p}\rangle$.

The time-dependent function $h(t)$ determines when the system and the detector are coupled.
If the full protocol occurs between time $t=0$ and $t=T$, we take  $h(t) = \delta(T-t) - \delta(t_1)$ where $\delta(t)$ is the Dirac delta and $0 < t_1< T$.
By using the Dirac delta, we model the fact that the system detector interaction occurs over much smaller time scales than all the other evolutions so that during the coupling we can consider the system dynamics ``frozen".
Under this assumption, the system-detector evolution is associated with the unitary evolution $U_\pm = \exp\{\pm i \lambda \hat{p} \otimes \hat{M} \}$.

In the intervals between the couplings with the detector, the system evolves with unitary evolution $U_i$ ($i=1, 2$) so that the total (system$+$detector) evolution is
\begin{equation}
	U_{tot} =  e^{ i \lambda \hat{p} \otimes \hat{M} } U_2 e^{-i \lambda \hat{p} \otimes \hat{M} } U_1
	\label{eq:U_tot}
\end{equation} 
where the explicit form of the $U_i$ will be determined below.

We consider as initial state $\lvert\psi_0\rangle = \lvert0\rangle \left( \frac{1}{\sqrt{N}} \sum \lvert{p}_D\rangle \right)$ where the sum is over $N$ detector states and the index $D$ denotes the states in the detector space.
The final state is $\ket{\psi_f} = U_{tot} \ket{\psi_0}$ associated to the density operator $\rho_{tot} = \ketbra{\psi_f}{\psi_f}$.
The final density matrix of the detector is $\rho_D^f = \Trace_S [\rho_{tot}] $ where $\Trace_S$ denotes the trace over the system degrees of freedom.

We define the quasi-characteristic function as \cite{solinas2015fulldistribution,solinas2016probing}
\begin{equation}
	\GFunc = \frac{_D{\matrixel{\bar{p}}{\rho_D^f }{-\bar{p}}}{_D}}
	{
	_D{\matrixel{\bar{p}}{\rho_D^0 }{-\bar{p}}}{_D}
    }.
	\label{eq:G_lambda_def}
\end{equation}
where $\ket{\pm \bar{p}}_D$ is a specific eigenstates of $\hat{p}$.
From a physical point of view, the quasi-characteristic function is the phase accumulated between the states $\ket{\pm \bar{p}}_D$ of the detector during the evolution \cite{solinas2015fulldistribution,solinas2016probing}.
This can be directly measured with interferometric techniques.

%The denominator of the above function is associated with the initial phase of the detector, i.e., we take the reduced detector matrix and calculate the off-diagonal term between the states $\ket{\pm p}$. In the specific case, $\matrixel{\bar{p}}{\rho_D^0 }{-\bar{p}}= 1/N$.
%Analogously, the numerator is associated with the final phase of the detector.
%The ratio between the two is therefore the phase accumulated between the states $\ket{\pm p}$ during the evolution.

More explicitly, we can now calculate the numerator of Eq.~(\ref{eq:G_lambda_def}). 
Since $U_i$ does not act on the detector state and $\ket{p}_D$ are eigenstates of $\hat{p}$, we have
\begin{equation}
\begin{split}
	_D\braket{\bar{p}\lvert \psi_f} =& \frac{1}{\sqrt{N}} \sum~_D\bra{\bar{p} } e^{ i \lambda \hat{p} \otimes \hat{M} } U_2 e^{-i \lambda \hat{p} \otimes \hat{M} } U_1 \ket{0} \ket{p}_D\\
	=& 	\frac{1}{\sqrt{N}} \sum \delta_{\bar{p} ,p } e^{ i \lambda p  \hat{M} } U_2 e^{-i \lambda p \hat{M} } U_1 \ket{0}\\
	=&\frac{1}{\sqrt{N}}  e^{ i \lambda\bar{p}  \hat{M} } U_2 e^{-i \lambda \bar{p} \hat{M}} U_1 \ket{0}.
\end{split}
\end{equation}
Analogously,
\begin{equation}
    \braket{\psi_f\lvert -\bar{p} }_D = \frac{1}{\sqrt{N}} \bra{0} U_1^\dagger e^{-i \lambda \bar{p}  \hat{M} } U_2^\dagger e^{i \lambda \bar{p} \hat{M} } .
\end{equation}
Noticing that $_D{\matrixel{\bar{p}}{\rho_D^f }{-\bar{p}}}{_D} = \Trace_S [ _D{\matrixel{\bar{p}}{\rho_{tot} }{-\bar{p}}}{_D} ]$, we have
\begin{eqnarray}
	\Trace_S [ _D{\matrixel{\bar{p}}{\rho_{tot} }{-\bar{p}}}{_D} ] 
	%&=& \frac{1}{N} \Trace_S [ e^{ i \lambda \bar{p}  \hat{M} } U_2 e^{-i \lambda \bar{p} \hat{M} } U_1 \ketbra{0}{0}  U_1^\dagger e^{- i \lambda \bar{p}  \hat{M} } U_2^\dagger e^{-i \lambda \bar{p} \hat{M} }] \nonumber \\
	&=& \frac{1}{N} \Trace_S [ \mU_\lambda \rho_S^0 \mU^\dagger_{-\lambda}]
\end{eqnarray}
where we have defined the operator $\mU_\lambda$ as 
\begin{equation}
	\mU_\lambda = e^{ i \lambda \bar{p}  \hat{M} } U_2 e^{-i \lambda \bar{p} \hat{M} } U_1.
	\label{eq:U_lambda}
\end{equation}
Finally, we can rewrite the quasi-characteristic function as  \cite{solinas2015fulldistribution,solinas2016probing}
\begin{equation}
	\GFunc = \Trace_S [ \mU_\lambda \rho_S^0 \mU^\dagger_{-\lambda}].
	\label{eq:G_lambda}
\end{equation}

In statistics, the knowledge of the characteristic function gives access to all the moments of the distribution since they are related to its derivative calculated in $\lambda$.
Since we are dealing with a quantum system, we have a quasi-characteristic function and the situation is more tricky since the associate probability distribution is not positively defined \cite{clerk2011full, solinas2015fulldistribution,solinas2016probing,SolinasPRA2017, solinas2021, solinas2022}, i.e., it is a quasi-probability distribution.
This makes it difficult to assign a precise meaning to the high moments.
However, it can be shown that the first moment is not affected by these problems, it has a practical and operative interpretation in terms of observable measurements  \cite{solinas2015fulldistribution,solinas2016probing}.

Given this premise, we are interested in the derivative
\begin{equation}
	-i \partial_\lambda \GFunc \Big \lvert_{\lambda=0} = -i\,\Trace_S [ \partial_\lambda \mU_\lambda \rho_S^0 \mU^\dagger_{-\lambda}
	+ \mU_\lambda \rho_S^0 \partial_\lambda \mU^\dagger_{-\lambda} 
	] \Big\lvert_{\lambda=0}.
	\label{eq:dG_lambda}
\end{equation}
This can be numerically obtained once the behavior of $\GFunc$ is known close to $\lambda =0$.

%%%%%%%%%%%%%%%%%%%%%%%%%%%%%%%%%%%%%%%%%%%%
\begin{figure}
    \begin{center}
    \includegraphics[scale=.4]{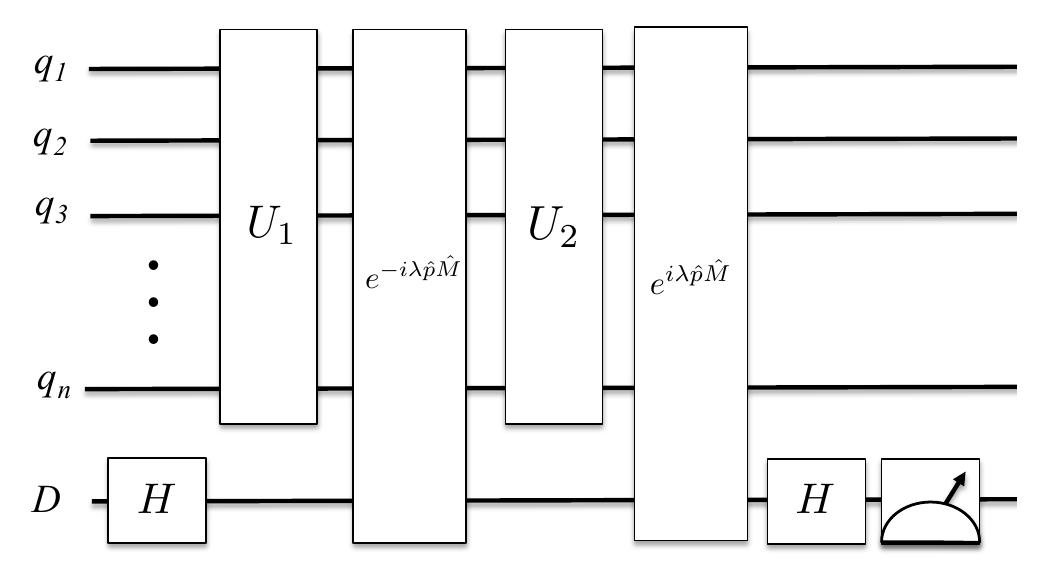}
       \end{center}
    \caption{
Scheme to perform a measure of the gradient using an ancilla qubit as a quantum detector.
The final Hadamard operation followed by the measurement corresponds to the measurement of the off-diagonal element of the density matrix and gives information about the accumulated phase.
} 
    \label{fig:nondem_protocol}
\end{figure} 
%%%%%%%%%%%%%%%%%%%%%%%%%%%%%%%%%%%%%%%%%%%%
From definition~(\ref{eq:U_lambda}) we can calculate $\mU^\dagger_{-\lambda}$ and the corresponding derivatives calculated in $\lambda =0$: $\partial_\lambda \mU_{\lambda} \Big\lvert_{\lambda=0}$ and  $\partial_\lambda \mU^\dagger_{-\lambda} \Big\lvert_{\lambda=0}$. We obtain
\begin{eqnarray}
	\partial_\lambda \mU_{\lambda} \Big\lvert_{\lambda=0} &=& i \bar{p} (\hat{M} U_2 U_1 - U_2 \hat{M} U_1) \nonumber \\
	\partial_\lambda \mU^\dagger_{-\lambda} \Big\lvert_{\lambda=0} &=& i \bar{p} (U^\dagger_1 U^\dagger_2 \hat{M} - U^\dagger_1 \hat{M} U^\dagger_2)
\end{eqnarray}
and, after algebraic manipulation using the cyclic property of the trace, the first derivative of $\GFunc$ reads
\begin{equation}
	-i \partial_\lambda \GFunc \Big\lvert_{\lambda=0} = 2 \bar{p} \Trace_S [U^\dagger_1 U^\dagger_2 \hat{M} U_2 U_1 \rho_S^0 - U^\dagger_1 \hat{M} U_1 \rho_S^0].
	\label{eq:first_derivative}
\end{equation}
The $2 \bar{p}$ can be taken equal to $1$ with opportune rescaling of $\hat{p}$.

Notice that in Eq.~(\ref{eq:first_derivative}) we have the difference of the average value of $\hat{M}$ evolved with $U_2 U_1$ and the one evolved with $U_1$.
By taking $U_1 = U(\vec \theta -s \ej{1} )$ and $U_2 U_1 = U(\vec \theta +s \ej{1} )$, using $f(\vec \theta)$ definition in Eq.~(\ref{eq:f_def}), we recover the numerator in Eq.~(\ref{eq:f_grad_def}).
In the $\vec \theta$ parameter space, $U_1$ is the transformation or the circuits that evolves the states to $\vec \theta -s \ej{1} $
 and $U_2 U_1$ generates the evolution from $\vec \theta -s \ej{1} $ to $\vec \theta +s \ej{1} $ (see Appendix~\ref{app:U_evolution}).

%We recall that $U_1$ is the transformation that maps the initial state $\ket{0}$ to the state $\ket{\psi({\vec \theta} -s)}$; so, it should be denoted as $U_1 = U({\vec \theta} -s,0)$

\subsection{Practical Implementation}
\label{sec:implementation}

A practical implementation of the protocol is shown in Fig.~\ref{fig:nondem_protocol}.
We need $n$ logical qubits plus one detector/ancilla qubit. The latter is initialised to $\ket{\eta_D^0} = (\ket{0} + \ket{1})/\sqrt{2}$.
This means that in the following $\hat{p} = \sigma_z$ (where $\sigma_z$ is the usual Pauli operator).
However, in the following, to distinguish the detector qubit from the logical ones, we keep using the $\hat{p}$ notation.

To calculate the gradient close to a value $\theta$ we need to choose $\lambda \ll 1$ (see Appendix~\ref{app:qubit_detector}).

The protocol to estimate the gradient of $f$ in the point $\vec \theta $ is the following 
\begin{enumerate}
\item Fix $\vec \theta$ and $\lambda \ll 1$.\\
	\begin{enumerate}
 	      \item apply a Hadamard operator to the detector qubit
		\item run the circuit $U({\vec \theta} -s \ej{1})$
		\item couple the system and the detector with the transformation $e^{- i \lambda \hat{p} \otimes \hat{M}}$
		\item run the circuits from ${\vec \theta} -s \ej{1}$ to ${\vec \theta} +s \ej{1}$: $U({\vec \theta} +s \ej{1}, {\vec \theta} -s \ej{1})$
		\item couple the system and the detector with the transformation $e^{ i \lambda \hat{p} \otimes \hat{M}}$
		\item apply a Hadamard operator to the detector qubit
		\item measure the detector qubit\\
	\end{enumerate}
\item repeat $m$ times to determine the detector population.
%\item change $\theta$ according to the optimisation method.
\end{enumerate}

The detector phase can be measured with interferometric or tomographic techniques.
%It can be shown that the measured population of the detector is related to the accumulated phase $\phi(\lambda)$ by the relations (Appendix~\ref{app:qubit_detector})
%\begin{eqnarray}
%	P_0 &=& \frac{1}{2} (1+ \cos^2 \phi) = \cos^2\frac{\phi}{2} \nonumber \\
%	P_1 &=& \frac{1}{2} (1- \cos^2 \phi)  = \sin^2\frac{\phi}{2} \nonumber 
%\end{eqnarray}
%
%The error on the population scales with the number of repetitions $m$ as $1/\sqrt{m}$.
%For small error in the phase $\delta \phi$ we obtain 
%\begin{equation}
%	\lvert\delta \phi\lvert = \frac{2}{\sqrt{m} \sin \left (2 \phi (\lambda) \right)}.
%\end{equation}
The gradient of $f$ estimated with the QNDM approach can be used to determine the updated vector parameter according to the optimizer algorithm.

%%%%%%%%%%%%%%%%%%%%%%%%%%%%%%%%%%%%%%%%%%%%%%%%%%%%%%%%%%%%%%%%%%%%%%

\section{Example and resources estimate}
\label{sec:example_and_resources}

To better understand the QNDM approach we present an example that is general enough to have many practical applications. For the sake of discussion, suppose we want to use a quantum computer to calculate the minimum average of a complex operator $\hat{M}$ in a high-dimensional quantum system.
The $\hat{M}$ operator can be written as the sum of weighted contributions \cite{McArdle2020} 
\begin{equation}
	\hat{M} = \sum_{j=1}^{J} h_j  \hat{P}_j =\sum_{j=1}^J h_j \prod_{i=1}^I \sigma_i^j
	\label{eq:Hamiltonian}
\end{equation}
where $h_j$ is a real coefficient, $\sigma_i^j$ is one of the Pauli operators where $i$ is the qubit index and $j$ denotes the term in the Hamiltonian \cite{McArdle2020} . Concurrently, $I$ and $J$ are two natural numbers such that we can consider $\hat{M}$ a complex operator and the $i$ ($j$) index runs from $1$ to $I$ ($J$).
The operators $\hat P_j = \prod_i \sigma_i^j$ are usually called Pauli string operators \cite{McArdle2020}.

These kinds of Hamiltonians describe a large number of different physical systems from condensed matter to quantum chemistry \cite{Wecker2015, McArdle2020}.
Before proceeding, two additional remarks must be made. First, the length of a Pauli string is bounded by the number of qubits $n$. Second, in physical systems because of the nature of the interaction between particles, usually, we have a maximum of four terms composing the Pauli strings, i.e., $I_{max} =4$ \cite{Wecker2015, McArdle2020}.

%%%%%%%%%%%%%%%%%%%%%%%%%%%%%%%%%%%%%%%%%%%%
\begin{figure}
    \begin{center}
    \includegraphics[scale=.55]{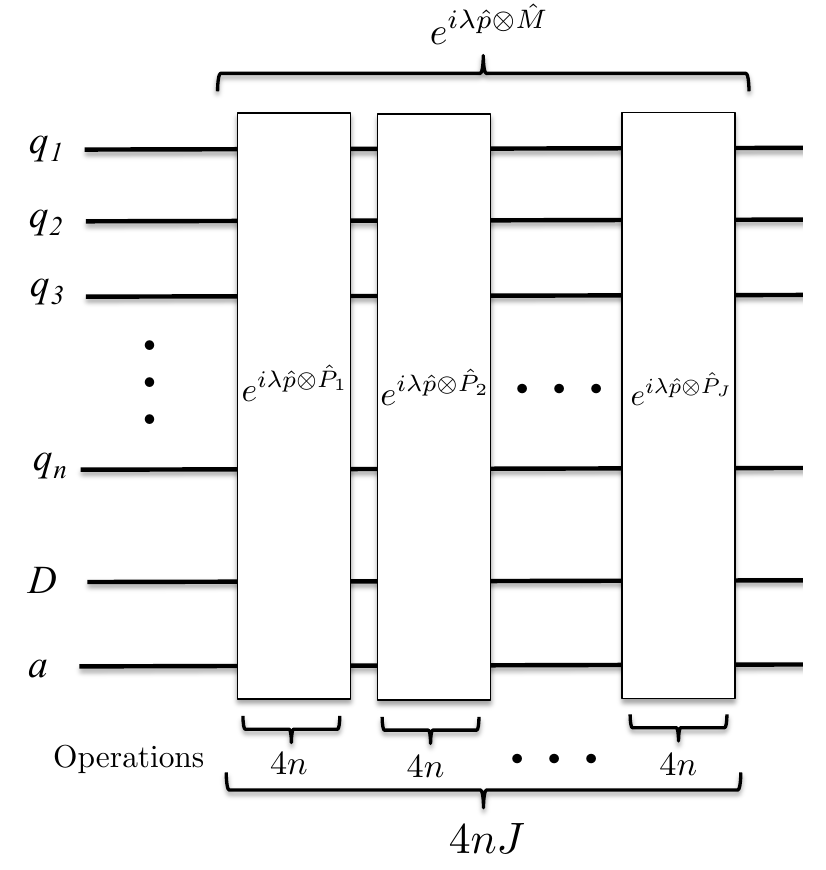}
       \end{center}
    \caption{The implementation of the exponential operator $\exp \{ i \lambda \hat{p} \otimes \hat{M} \}$ in a quantum computer with an additional ancilla qubit $a$ (see Appendix~\ref{app:P_j_implmentation} for details).
    The operator is decomposed in a product of Pauli string operator $\exp \{ i \lambda \hat{p} \otimes  \hat P_j \}$.
    The number of elementary operations needed for every operator is shown below.
    The total number of the elementary operators is $4nJ$ where $n$ is the number of logical qubits and $J$ is the number of Pauli strings in $\hat{M}$.
 } 
    \label{fig:P_j_product}
\end{figure} 
%%%%%%%%%%%%%%%%%%%%%%%%%%%%%%%%%%%%%%%%%%%%%%%%%%%%%%%%%%%%%%%%%%%%%%%%%%%%%%%%%%%%%%%%

If $\hat{M}$ is a complex operator, the implementation of the exponential coupling $\exp \{ i \lambda \bar{p}  \hat{M} \}$ needed for the QNDM is, in general, very resource-consuming.
This might seem a critical drawback, but the QNDM approach offers a clear and elegant way to bypass this problem.
For the purpose of measuring the average of the gradient in the QNDM approach, instead of implementing the operator $\exp{\{ i \lambda \hat p \otimes  \sum_j h_j \hat P_j \} }$, we can implement $\prod_j \exp{\{ i \lambda h_j \hat p \otimes \hat P_j \}} $ as shown in Fig.~\ref{fig:P_j_product}.

To clarify this point, let us consider the operator $\hat{M}= h_1 \hat{P}_1 + h_2 \hat{P}_2 + h_3 \hat{P}_3$.
If we implement the operator $U_{full}= \exp{\{ i \lambda \hat p \otimes \hat M \} }$ and then take the derivative with respect to $\lambda$ to obtain $\GFunc$, we obtain
\begin{equation}
 i \partial_ \lambda U_{full} \Big\lvert_{\lambda = 0 } = \bar{p} (h_1 \hat{P}_1 + h_2 \hat{P}_2 + h_3 \hat{P}_3).
\end{equation}

If we implement the operator  $U_{simpl}= \prod_j U_j  =  \exp{\{ i \lambda  h_1 \hat p \otimes \hat{P}_1 \}} \exp{\{ i \lambda  h_2 \hat p \otimes \hat{P}_2 \}} \exp{\{ i \lambda  h_3 \hat p \otimes \hat{P}_3 \}}$ (with $U_j = \exp{\{ i \lambda h_j \hat p \otimes \hat{P}_j \}}$), the derivative reads
\begin{equation}
\begin{split}
 i \partial_ \lambda U_{simpl} \Big\lvert_{\lambda = 0 } =& \bar{p} \Big[  (h_1 \hat{P}_1) U_1 U_2 U_3 +  \\
 +& U_1  (h_2 \hat{P}_2) U_2 U_3+ \\
 +& U_1 U_2 U_3 (h_3 \hat{P}_3) \Big] \Big\lvert_{\lambda = 0 } \\
 =& \bar{p} (h_1 \hat{P}_1 + h_2 \hat{P}_2 + h_3 \hat{P}_3).
 \end{split}
\end{equation}
Therefore, despite being different, the unitary operators $U_{full}$ and $U_{simpl}$ contribute in the same way to the averages.
The main advantage of this approach lies in the fact that the single Pauli string operator $\hat{P}_j$ can be more easily implemented in a quantum computer in terms of single qubit operators.

In Fig.~\ref{fig:P_j_product}, we show how the product of exponential can be implemented. 
The additional {\it ancilla} qubit is needed to properly implement the $\exp{\{ i \lambda h_j \hat p \otimes \hat P_j \}}$ operator.
The technical details can be found in Appendix~\ref{app:P_j_implmentation} where we also discuss the resources needed in terms of elementary logical quantum gates.
It turns out that every $\exp \{ i \lambda \hat{p}  P_j \}$ block needs a maximum of $4n$ logical operation (Appendix~\ref{app:P_j_implmentation}).
If $\hat{M}$ is composed of $J$ Pauli strings we need a maximum of $4nJ$ logical operations to implement the full $\exp \{ i \lambda \hat{p}  \hat M \}$ operator.

At this point, we can estimate the cost of the circuit for the QNDM in Fig.~\ref{fig:nondem_protocol}.
Once we have implemented the operator $U_1 = U(\vec \theta -s \ej{1} )$ with $k$ logical operators, usually, we need a negligible number of logical gates to arrive at the point $\vec \theta +s \ej{1}$. 
In fact, the shift along the $\ej{1}$ direction involves only a single qubit rotation as in Eq. (\ref{eq:U_def}).
%to implement $U_2 U_1 = U(\vec \theta +s \ej{1} )$.
Therefore, the total cost of $U_1$ and $U_2$ operators is approximately $k$.
The two system-detector couplings $\exp \{ \pm i \lambda \bar{p}  \hat M \}$ are performed with $8 n J$ elementary operations. 
The total cost of the QNDM circuit in Fig.~\ref{fig:nondem_protocol} is $8nJ + k$.

Since the circuit must be repeated $m$ times to have the desired statistical accuracy, the total cost to estimate the gradient of a quantum operator is of $(k+ 8nJ)m$ elementary logical operators.

As an additional side remark, we want to emphasize that this result does not rely on any approximation differently from the usual Lie-Trotter-Suzuki decomposition \cite{McArdle2020}. The latter depends on the assumption that $\lambda p h_j \ll 1$.
This implies that it might be necessary to iterate the process to implement the desired operator for any $\lambda p h_j$.
On the contrary, the present methods can be used for any value of $\lambda p h_j$, although a small value of $\lambda$ might be helpful for phase evaluation.

\section{Second and Higher derivatives}
\label{sec:higher_derivatives}

\begin{figure}
    \begin{center}
    \includegraphics[scale=.6]{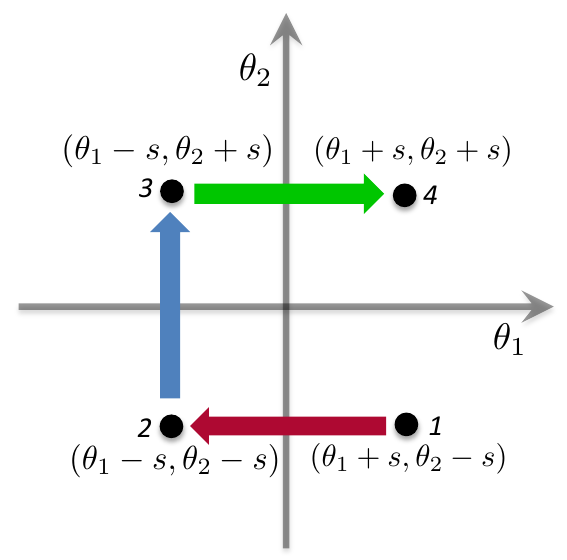}
       \end{center}
    \caption{
Path in the bidimensional $(\theta_1, \theta_2 )$ parameter space of the unitary operator $U(\vec \theta)$ to obtain the second derivative.
First, we implement $ U\big(\vec \theta - s (\ej{1}+\ej{2}) \big)$ (point $(\theta_1 -s, \theta_2 - s)$), the implement the shift of $2s$ along the $\theta_1$ coordinate to arrive to point $2$, i.e., $(\theta_1 + s, \theta_2 + s)$.
Then, we move to point $3$, i.e., $(\theta_1 - s, \theta_2 + s)$ and, finally, to point $4$, $(\theta_1 + s, \theta_2 - s)$.
Notice that the path is not unique. In fact, one can equivalently go around in the other direction (i.e. points 1, 4, 3 and 2) or start from point 3 (both directions).
} 
    \label{fig:parameter_path}
\end{figure}

The QNDM is even more convenient if we want to calculate the second derivative in Eq.~(\ref{eq:second_derivative}).
In this case, we just need to couple the system and the detector four times in a single run.
The unitary transformation we implement is 
\begin{equation}
	\mU_\lambda = e^{ i \lambda \bar{p}  \hat{M} } U_4 e^{-i \lambda \bar{p} \hat{M} } U_3 e^{ i \lambda \bar{p}  \hat{M} } U_2 e^{-i \lambda \bar{p} \hat{M} } U_1
	\label{eq:U_lambda_second_order}
\end{equation}
with the transformations (see Appendix~\ref{app:U_evolution})
\begin{eqnarray}
	U_1 &=& U\big(\vec \theta + s (\ej{1}-\ej{2}), 0 \big) \nonumber \\
	U_2 &=& U\big(\vec \theta - s (\ej{1}+\ej{2}), \vec \theta + s (\ej{1}-\ej{2}) \big) \nonumber \\
	U_3 &=& U\big(\vec \theta + s (-\ej{1}+\ej{2}),\vec  \theta - s (\ej{1}+\ej{2} ) \big)\nonumber \\
	U_4 &=& U\big(\vec \theta + s (\ej{1}+\ej{2}), \vec \theta + s (-\ej{1}+\ej{2}) \big)
	\label{eq:U_operators_2D}
\end{eqnarray}
where we have denoted with $U(\vec \theta_i , \vec \theta_j)$ the operator that allows us to pass from $\vec \theta_j$ to  $\vec \theta_i$ in the $\vec \theta$ parameter space (Appendix~\ref{app:U_evolution}). 
In Figure~\ref{fig:parameter_path} the process is shown in a bidimensional space, i.e., $\vec \theta = (\theta_1, \theta_2)$.
Formally the $\GFunc$ and its derivative are written as in Eqs.~(\ref{eq:G_lambda}) and~(\ref{eq:dG_lambda}).
By direct calculation, we obtain (see Appendix~\ref{app:U_evolution})
\begin{equation}
\begin{split}
	-i \partial_\lambda \GFunc \Big \lvert_{\lambda=0} 
%	&=& \Trace_S [ U^\dagger (\vec \theta + s (\ej{1}+\ej{2})) M U (\vec \theta + s (\ej{1}+\ej{2})) \rho_S^0  \nonumber \\
%	&&-U^\dagger (\vec \theta + s (-\ej{1}+\ej{2})) M U (\vec \theta + s (-\ej{1}+\ej{2})) \rho_S^0 \nonumber \\	
%	&&+U^\dagger (\vec \theta + s (\ej{1}-\ej{2})) M U (\vec \theta + s (\ej{1}-\ej{2})) \rho_S^0 \nonumber \\
%	&&-U^\dagger (\vec \theta - s (\ej{1}+\ej{2})) M U (\vec \theta - s (\ej{1}+\ej{2})) \rho_S^0 
%	] \nonumber \\
	&= f(\vec \theta +s (\ej{1}+\ej{2})) +\\
        &- f(\vec \theta +s (-\ej{1}+\ej{2}))+\\
        &-f(\vec \theta +s (\ej{1}-\ej{2}))+ \\
        &+ f(\vec \theta -s (\ej{1}+\ej{2})).
 \end{split}
\end{equation}
By dividing this by $2 \sin^2 s$ we obtain the expression in Eq.~(\ref{eq:second_derivative}). 
%
%%%%%%%%%%%%%%%%%%%%%%%%%%%%%%%%%%%%%%%%%%%%
%%%%%%%%%%%%%%%%%%%%%%%%%%%%%%%%%%%%%%%%%%%%%%%%%%%%%%%%%%%%%%%%%%%%%%%%%%%%%%%%%%%%%%%%
%
Interestingly, this can be evaluated with 
\begin{itemize}
\item $n$ qubits
\item {\it a single circuit} of approximate $k$ the logical gates (needed for $U (\vec \theta + s (\ej{1}+\ej{2}))$) 
\item As above, we are measuring the detector phase and we need only $m$ repetitions to have an error of $1/\sqrt{m}$.
\end{itemize}
Notice that in the DM approach the measure of the second derivative needs the estimate of the average of four observables, i.e., 
$M(\vec \theta + s (\pm \ej{1}\pm\ej{2}))$.
On the contrary, with the QNDM we obtain the same information with a single-phase estimation.
The same scheme can be used to extract information about the derivative of arbitrary order.
Interestingly, the increase in the derivative order implies an increase in the resources (logical gate and measurements) for the DM while  only weakly affecting the complexity of the QNDM.
Following the gate counting procedure discussed in Sec.~\ref{sec:example_and_resources}, we estimate that to estimate the gradient of $f$ with the QNDM, we need a maximum of $(k+ 16 nJ)m$ elementary logical operators.

\section{Comparison with other approaches}
\label{sec:advantages}

We are now in a position to discuss the advantages of the QNDM with respect to the usual DM protocol \cite{Kandala2017, Havlicek2019, Mari2021}.

We must stress that the cost of the resources (in terms of circuit repetitions, number of unitary operations used and so on) depends critically on the specific problem we want to solve (see, for example, Ref. \cite{Wecker2015}).
A second factor to be kept into account in the comparison is the quantum hardware used. 
Usually, the approaches that use the quantum phase to store the information (this article and \cite{OBrien2019, McArdle2020, Abrams1999, Banchi2021}) need efficient hardware, error mitigation and correction techniques since they must preserve quantum coherences for long times.
Therefore, for noisy hardware, the quantum phase approaches can be less efficient than the direct measure ones \cite{McArdle2020, Mari2021}.

Given these premises, we are interested in giving a rough estimate of the advantages and pointing out where these can be found without entering the details of the simulated physical system.
In addition, we suppose that the noise can be eliminated so that the dephasing effect can be neglected.

\subsection{Gradient and second derivative measurement}
We suppose that we need $\approx k$ logical operators to implement both $U(\vec \theta \pm s \ej{1})$.
The more straightforward way to evaluate the function $f(\vec \theta)$ is to run the circuit/dynamics and then measure the operator $\hat{M}$ \cite{Kandala2017, Havlicek2019, Mari2021}.

As discussed above, since the complexity of the system usually increases with its dimension, for realistic
and practical situations this can be an extremely difficult task \cite{McArdle2020,Verteletskyi2020, Yen2020,Yen2021}. Therefore, we need to separately measure every single Pauli string $\hat{P}_j$.

The procedure is the following.
First, we run the circuit with the unitary operator $U(\vec \theta - s \ej{1})$ and measure $\hat{P}_j$.
Repeating this protocol $m$ times we can obtain the average value $\average{\hat{P}_j(\vec \theta - s \ej{1})}$ with precision scaling as $1/\sqrt{m}$.
As discussed in the appendix~\ref{app:P_j_implmentation}, since the measure in a quantum computer are always performed in the standard basis, i.e., eigenstates of $\sigma_z$, a generic Pauli string we have to rotate a maximum of $n$ qubit with Hadamard gates.
For example, to measure the Pauli string $\sigma_x^1 \sigma_x^2 \sigma_x^3$ we need to apply the Hadamard operators $H_{had}^1 H_{had}^2 H_{had}^3$.

We repeat the procedure for all the $\hat{P}_j$ for a total of $J$ repetitions.
By summing the weighted averages we obtain 
$\average{\hat{M}(\vec \theta - s \ej{1})} = \sum_j h_j \average{\hat{P}_j(\vec \theta - s \ej{1})} = f(\vec \theta - s \ej{1})$.
Then, we repeat the same procedure $m J$ times running the circuit with the unitary operator $U(\vec \theta + s \ej{1})$ and measure $\hat{M}$ to obtain $\average{M(\vec \theta + s \ej{1})} = f(\vec \theta + s \ej{1})$.

In terms of resources, summing all the contributions we must apply $2 m J (k+n) $.

For the second derivative, the counting is analogous but we have to consider that we need four averages of the $\hat M$ operators and, this, a total of $4 m J (k+n) $.

%%%%%%%%%%%%%%%%%%%%%%%%%%%%%%%%%%%%%%%%%%%%
\begin{center}
\begin{table}[t]
  \begin{tabular}{ | l | c | c | c | c| }
    \hline
   			& {\it qubits} & {\it iterations} & {\it gates}  \\ 
				      \hline
$\partial_1 f$--DM  & $n$                          & $2 m$         & $\approx 2 m J (k+2n)$    \\ 
					\hline
$\partial_2 f$--DM  & $n$                          & $4 m$         & $\approx 4 m J (k+2n)$    \\ 
					\hline
$\partial_1 f$--QNDM          & $n+2$                          & $m$          & $ \approx m (k + 8 n J)$          \\ 
				\hline
$\partial_2 f$--QNDM          & $n+2$                          & $m$          & $\approx m (k + 16 n J)$          \\ 
				\hline

  \end{tabular}
\caption{Comparison between different methods to measure the gradient  and second derivative of $f(\vec \theta)$: Direct Measurement (DM) and  Quantum Non-Demolition Measurement (QNDM). 
The parameter $m$ determines the error on the estimate of the gradient. The parameter $k$ determines the number of gate operations needed to implement the operators $U(\vec \theta)$. Since the shift $s$ occurs along a single direction $\ej{1}$, we suppose that the number of gates to implement $U(\vec \theta \pm s \ej{1})$. For small $s$ is approximatively the same and equal to $k$.}
\label{tab:highertable}
  \end{table}
\end{center}

This resource counting must be compared to the one estimate for the QNDM in Sec.~\ref{sec:gradient} and~\ref{sec:higher_derivatives}.
Table~\ref{tab:highertable} summarizes the resources needed for the two approaches.
As can be seen, the QNDM has a clear advantage in the number of iterations of the circuit. This is because in a single step we can store information about two average values of $\hat M$, i.e., about the gradient.

We have to stress that in the QNDM approach we usually run a more complex and deeper quantum circuit. 
For a comparison of the logical quantum gates, we can refer to Table~\ref{tab:highertable}.
We can consider two limits.
First, we notice that the number of logical gates $k$ needed to apply the unitary operator $U(\vec \theta)$ is usually greater than $n$ \cite{Wecker2015}.
In the limit of $k \gg n J$, we have that the resource ratio scales as ${\rm DM}/{\rm QNDM} \approx J$.
Therefore, depending on the number of Pauli strings to be measured, we can have a substantial advantage with the QNDM.
This regime is likely to be interesting for the simulation in quantum chemistry.
For example, for the simulations of medium complex molecules, we could estimate \cite{Wecker2015}:   $k\approx 10^9-10^{10}$ operations to implement $U(\vec \theta)$, $n \approx 10^2-10^3$ (that is the dimension of quantum computer available in the next years) and $J > 10^3$ number of Pauli string in the Hamiltonian. With these numbers, $k \gg n J$ and we have a reduction of roughly $J$ logical operations.

In the opposite limit $k \ll n J$, we have ${\rm DM}/{\rm QNDM} \approx k/n$.
Since the number of operations to implement $U(\vec \theta)$ is greater than the number of qubit $n$, also, in this case, the QNDM has an advantage with respect to the DM.

This advantage increases with the order of the derivative we need.
While the number of repetitions of the DM circuit increases, for the QNDM is constant.
In most common approaches and optimizers, only the first and second derivatives of the cost function are usually exploited.
However, there are some indications that derivatives of a high order can be used to treat classically hard optimization problems \cite{ahookhosh2021A,ahookhosh2021B}.
In addition, in some specific problems in finance, the derivative themselves carry the relevant information \cite{Stamatopoulos2022}.
In these cases, the possibility to efficiently measure the high-order derivatives of the cost function with the QNDM can give substantial advantages.

\section{Conclusion}\label{sec:conclusions}

We have discussed an alternative approach to estimate the gradient and higher-order derivatives of a cost function on quantum hardware by using a quantum detector.
The relevant information is stored in the phase of the detector.
Since the evolution of the system is not affected by the detector, this is usually referred to as Quantum Non-Demolition Measurement \cite{clerk2011full}.
At its roots, the idea is related to the weak values and weak measurements in quantum mechanics \cite{Aharonov1988weak,bednorz2012nonclassical}.

The main advantage of the QNDM is that a single quantum circuit allows us to extract the information about the gradient, i.e., the first derivative, and higher derivative of the average values of a quantum observable.
This reduces the number of circuit iterations needed for the application of the most common optimization algorithms such as gradient descent and Newton optimization.

In terms of logical operation needed, the QNDM approach implies the use of more complex quantum circuits since we need to store information in the detector phase.
A quantitative comparison with the Direct Measurement approaches depends critically on the problem we want to solve.
However, by considering a fairly general operator structure [see Eq.~(\ref{eq:Hamiltonian})], we were able to extract some general trends.
The QNDM approach is usually advantageous in terms of logical operator needed.
This advantage increases in the cases in which the quantum operator is written as the sum of many Pauli strings.
As above, for higher derivatives, there is a further advantage in the reduction of logical operators.

% The QNDM gives three main advantages.
% First, fixed the desired accuracy, it reduces the number of iterations of the protocol.
% Second, it reduces the number of the logical gates needed for the estimate derivatives of the cost function.
% These advantages with respect to the direct measurement approaches increase with the order of derivatives that we want to estimate.
% The third advantage consists in the possibility to easily extract information about complex quantum observables composed 
% by different non commuting contributions \cite{McArdle2020, Verteletskyi2020,Yen2020,Yen2021}.
% In the QNDM, all the information about the average of the observable is directly accessible with the detector phase measurement while in the DM approach several different measurements must be done to obtain the desired average observable.

The fast and accurate estimation of the gradient and high derivatives plays a key role in the variational quantum circuits \cite{McArdle2020, Cerezo2021}.
These can be applied to numerous problems in physics and chemistry as well as optimization problems.
The gradient estimation performed by a quantum computer is fed to a classical computer that processes the information to reach the minimum of the cost function.
This quantum-classical hybrid scheme has the potential to be one of the first practical cases in which we have a quantum advantage over classical computers.

In this perspective, any approach that decreases the quantum resources brings these applications closer to the actual implementation on a quantum computer.
Further studies are needed to fully gauge the impact and advantages of the QNDM approach.
However, especially for the NISQ computers where the resources are limited, the QNDM approach could offer a valuable alternative to reduce the resources and efficiently implement some variational quantum circuits.

\begin{acknowledgments}
The authors acknowledge financial support from INFN. This work has been carried out while Giovanni Minuto was enrolled in the Italian National Doctorate on Artificial Intelligence run by Sapienza University of Rome in collaboration with Dept. of Informatics, Bioengineering, Robotics, and Systems Engineering, Polytechnic School of Genoa.
\end{acknowledgments}

%%%%%%%%%%%%%%%%%%%%%%%%%%%%%%%%%%%%%%%%%%%%%%%%%%%%%%%%%%%%
%\appendix
%\setcounter{equation}{0}
%
%
%\pagebreak
%\widetext
%\begin{center}
%\textbf{\large Appendix}
%\end{center}
%%%%%%%%%%% Merge with supplemental materials %%%%%%%%%%
%%%%%%%%%%% Prefix a "S" to all equations, figures, tables and reset the counter %%%%%%%%%%
%\setcounter{equation}{0}
%\setcounter{figure}{0}
%\setcounter{table}{0}
%\setcounter{page}{1}
%\makeatletter
%\renewcommand{\theequation}{S\arabic{equation}}
%\renewcommand{\thefigure}{S\arabic{figure}}
%\renewcommand{\bibnumfmt}[1]{[S#1]}
%\renewcommand{\citenumfont}[1]{S#1}

\appendix
\setcounter{equation}{0}

\section{Unitary evolution in the parameter space}
\label{app:U_evolution}

In the main text, we have written the total unitary transformation to measure the gradient as $U_{tot} =  e^{ i \lambda \hat{p} \otimes \hat{M} } U_2 e^{-i \lambda \hat{p} \otimes \hat{M} } U_1$.
The $U_1$ and $U_2$ operation can be thought of as a sequence of non-commuting logical quantum gates or, alternatively using a more physical language, generated by a time-dependent Hamiltonian $H(\theta(t))$.

We consider the change in ${\vec \theta}$ along a single direction $\ej{1}$ so that the difference between the points ${\vec \theta} \pm s \ej{1}$ is parametrized only by $s$.
To build the operator $U({\vec \theta})$, we divide the transformation into $N$ steps as usually done for the evolution of a quantum system or the implementation in a quantum circuit.
Similarly, we suppose that to implement the operators  $U({\vec \theta} - s \ej{1})$ and $U({\vec \theta} + s \ej{1})$ we need $ N-N_1$ and $N+N_2$ operations, respectively.
Denoting the single unitary operation a $u_k$, we can write
\begin{equation}
 	U_1 = u_{N-N_1} ... u_{1} = U({\vec \theta} - s \ej{1})
\end{equation}
and $U_2 = \Pi_{j=1}^{N+N_2} u_j = \Pi_{j=N-N_1}^{N+N_2} u_j U_1 $.
%\begin{equation}
%	U_2 = u_{N+N_2} u_{N+N_2-1} ...  u_{N} u_{N-1} ... u_{N-N_1+1} u_{N-N_1} ... u_{1} = u_{N+N_2} u_{N+N_2-1} ...  u_{N} u_{N-1} ... u_{N-N_1+1}  U_1.
%\end{equation}
Therefore, we can obtain $U_2$ by implementing first $U_1$ and then applying other $N_2+N_1$ operators.

Using the mapping $U_1= U({\vec \theta} - s \ej{1})$ and $U_2= U({\vec \theta} + s \ej{1})$, we rewrite the above equation as 
\begin{equation}
	U({\vec \theta} + s \ej{1}) = U({\vec \theta} + s \ej{1}, {\vec \theta} - s \ej{1}) U({\vec \theta} - s \ej{1})
\end{equation}
where we have implicitly denoted with $U({\vec \theta} + s \ej{1}, {\vec \theta} - s \ej{1})$ the operator that allows us to pass from ${\vec \theta} - s \ej{1}$ to  ${\vec \theta} + s \ej{1}$ with $N_2+N_1$ operators.

In a similar way, we can implement the series of operators 
$\mU_\lambda = e^{ i \lambda \bar{p}  \hat{M} } U_4 e^{-i \lambda \bar{p} \hat{M} } U_3 e^{ i \lambda \bar{p}  \hat{M} } U_2 e^{-i \lambda \bar{p} \hat{M} } U_1$ to measure the second derivative.
The $U_j$ operators (with $j=1,..,4$) correspond to the operators working in the $\vec \theta$ space.
%First, we perform the transformation to the point in the parameter space ${\vec \theta} - s (\ej{1} + \ej{2})$.
%Then, we implement the tranformation ${\vec \theta} - s (\ej{1} + \ej{2}) \rightarrow	{\vec \theta} + s (\ej{1} + \ej{2})$.
%We go back to ${\vec \theta} $ and then implement the second ${\vec \theta} \rightarrow {\vec \theta} + s (-\ej{1} + \ej{2}) \rightarrow	{\vec \theta} + s (\ej{1} - \ej{2})$. From these we obtain the mapping discussed the main text.

We define the quasi-characteristic function as in Eq.~(\ref{eq:G_lambda_def}) and use the above expression for $\mU_\lambda$.
Taking the first derivative with respect to $\lambda$, and using the mapping 
\begin{eqnarray}
	U (\vec \theta + s (\ej{1}-\ej{2})) &=& U_1 \nonumber \\
	U (\vec \theta - s (\ej{1}+\ej{2})) &=& U_2 U_1\nonumber \\
	U (\vec \theta + s (-\ej{1}+\ej{2})) &=& U_3 U_2 U_1 \nonumber \\
	U (\vec \theta + s (\ej{1}+\ej{2})) &=& U_4 U_3 U_2 U_1,
\end{eqnarray}
we obtain
\begin{widetext}
\begin{equation}
\begin{split}
    -i \partial_\lambda \GFunc \Big\lvert_{\lambda=0} =& \Trace_S [U^\dagger (\vec \theta + s (\ej{1}+\ej{2})) M U (\vec \theta + s (\ej{1}+\ej{2})) \rho_S^0 -U^\dagger (\vec \theta + s (-\ej{1}+\ej{2})) M U (\vec \theta + s (-\ej{1}+\ej{2})) \rho_S^0 \\
    &-U^\dagger (\vec \theta + s (\ej{1}-\ej{2})) M U (\vec \theta + s (\ej{1}-\ej{2})) \rho_S^0 
	+U^\dagger (\vec \theta - s (\ej{1}+\ej{2})) M U (\vec \theta - s (\ej{1}+\ej{2})) \rho_S^0 ] \\
	=& f(\vec \theta +s (\ej{1}+\ej{2})) - f(\vec \theta +s (-\ej{1}+\ej{2}))
	-  f(\vec \theta +s (\ej{1}-\ej{2})) + f(\vec \theta -s (\ej{1}+\ej{2}))
\end{split}
\end{equation}
\end{widetext}
%
%\begin{equation}
%\begin{split}
%    -i \partial_\lambda \GFunc \Big\lvert_{\lambda=0} =& \Trace_S [U^\dagger (\vec \theta + s (\ej{1}+\ej{2})) M U (\vec \theta + s (\ej{1}+\ej{2})) \rho_S^0 +\\
%	-&U^\dagger (\vec \theta + s (-\ej{1}+\ej{2})) M U (\vec \theta + s (-\ej{1}+\ej{2})) \rho_S^0 +\\
%    -&U^\dagger (\vec \theta + s (\ej{1}-\ej{2})) M U (\vec \theta + s (\ej{1}-\ej{2})) \rho_S^0 +\\
%	+&U^\dagger (\vec \theta - s (\ej{1}+\ej{2})) M U (\vec \theta - s (\ej{1}+\ej{2})) \rho_S^0 ]= \\
%	=& f(\vec \theta +s (\ej{1}+\ej{2})) - f(\vec \theta +s (-\ej{1}+\ej{2}))\\ 
%	- & f(\vec \theta +s (\ej{1}-\ej{2})) + f(\vec \theta -s (\ej{1}+\ej{2}))
%\end{split}
%\end{equation}
where in the last line we have used the definition of $f(\vec \theta)$ in Eq.~(\ref{eq:f_def}).
A part from a $2 \sin s$ factor, this is the second derivative as in Eq.~(\ref{eq:second_derivative}).
By denoting with $U(\vec \theta_2 , \vec \theta_1)$ the operator that allows us to pass from $\vec \theta_1$ to  $\vec \theta_2$ , we obtain the explicit form of the $U_j$ operators as in Eq.~(\ref{eq:U_operators_2D}) of the main text.

\section{Detection with a qubit}
\label{app:qubit_detector}

We consider the case in which a qubit, i.e., a two-level system, is used as a detector.
The initial state of the detector is fixed at $\ket{\eta_0} = \cos \alpha/2 \ket{0} + \sin \alpha/2 \ket{1}$.
The initial contribution to $\GFunc$ is $\braket{0\lvert\eta_0} \braket{\eta_0\lvert1} = 1/2~\sin \alpha$.

After the evolution the detector acquires a phase $\phi(\lambda)$ so that its final state is $\ket{\eta_f} = \cos \alpha/2 \ket{0} + \sin \alpha/2 e^{- i \phi(\lambda)} \ket{1}$.
The numerator of the $\GFunc$ is $\braket{0\lvert\eta_f} \braket{\eta_f\lvert1} = 1/2~\sin \alpha e^{ i \phi(\lambda)}$ and 
\begin{equation}
	\GFunc = e^{i \phi(\lambda)}.
\end{equation}
is exactly the phase accumulated during the evolution.

We can verify that $\GFunc$ has the properties discussed in the previous section recalling that for $\lambda=0$ there is no system-detector interaction and, thus, no phase is accumulated, i.e., $\phi(0) = 0$.

In particular, $-i \partial_\lambda \GFunc \lvert_{\lambda =0 } =  \partial_\lambda \phi(\lambda)\lvert_{\lambda=0}$; therefore, the average value and the  gradient are related to the derivative of the accumulated phase evaluated in the origin.

In the experiments, we need to calculate $\partial_\lambda \phi(\lambda)\lvert_{\lambda=0}$ numerically.
We can approximate it as
\begin{equation}
	-i \partial_\lambda \GFunc \lvert_{\lambda =0 } = \partial_\lambda \phi(\lambda)\lvert_{\lambda=0} \approx \frac{\phi(\Delta \lambda) - \phi(0)}{\Delta \lambda} = 
	\frac{\phi(\Delta \lambda) }{\Delta \lambda}
\end{equation}
We can calculate the desired quantity by evaluating the protocol and the phase at a small $\Delta \lambda \ll 1$.

%\section{Error on the phase measurements}
%\label{app:phase_error}

%Let us consider the final detector state $\ket{\eta_f} = \cos \alpha \ket{0} + \sin \alpha e^{- i \phi(\lambda)} \ket{1}$.
%By applying a Hadamard gate and measuring we obtain the probabilities to measure $0$ and $1$
%\begin{eqnarray}
%	P_0 &=& (1+ \cos^2 \phi)\cos^2 \alpha= 2\cos^2 \frac{\phi}{2}\cos^2 \alpha \nonumber \\
%	P_1 &=& (1- \cos^2 \phi)\sin^2 \alpha=2\cos^2 \frac{\phi}{2}\sin^2 \alpha, \nonumber 
%\end{eqnarray}
%respectively.
%For the initial state with $\alpha = \pi/4$, we recover the results of the main text.
%%
%The statistical error on, say, $P_0$ scales as $ \delta P_0 = \beta/\sqrt{m}$ where $m$ is the number of repetitions and $\beta$ a proportionality constant.
%At the same time, if we consider how $P_0$ changes under a small $\delta \phi$ variation, we have that
%\begin{equation}
% \delta P_0 = -\cos \frac{\phi}{2} \sin \frac{\phi}{2} \delta \phi
%\end{equation}
%and, then, 
%\begin{equation}
%	\lvert\delta \phi\lvert = \frac{2 \beta}{\sqrt{m} \sin \phi}.
%\end{equation}

\section{Measuring the Pauli string - QNDM and DM circuit implementation}
\label{app:P_j_implmentation}

%%%%%%%%%%%%%%%%%%%%%%%%%%%%%%%%%%%%%%%%%%%%
\begin{figure}
    \begin{center}
    \includegraphics[scale=.6]{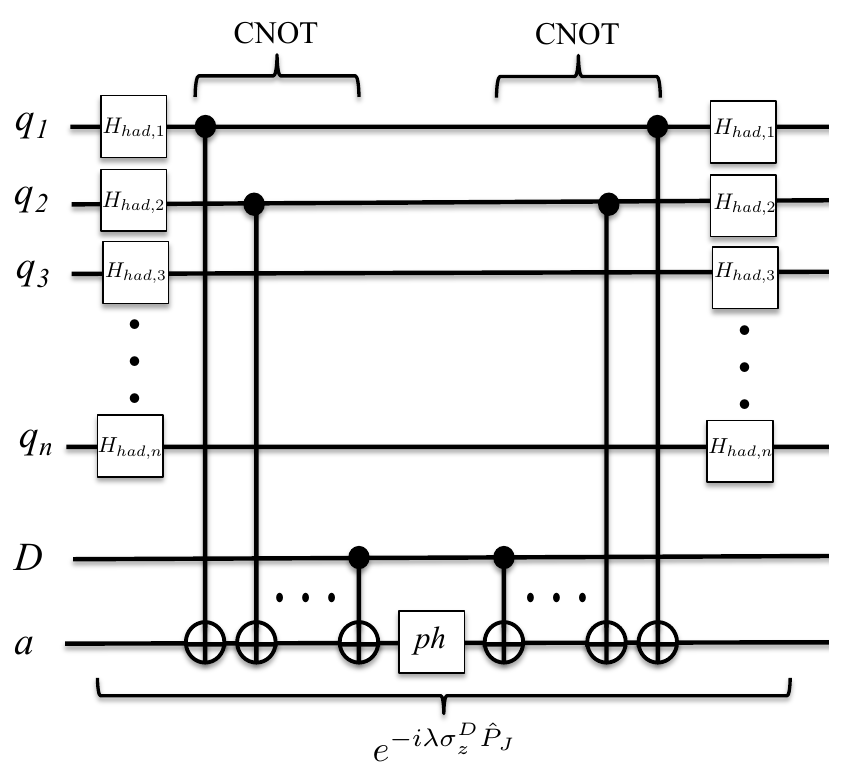}
       \end{center}
    \caption{The circuit to implement the operator $\exp \{ i \lambda \sigma_z^D  \prod_i \sigma_i^j \}$. We need a maximum of $2(n+1)$ Hadamard-like gates (to diagonalize the $\sigma_x^k$ or $\sigma_y^k$), $2(n+1)$ CNOT gates and a phase gate (denoted as $ph$) for a total of $\approx 4n$ elementary gates.
    Notice that if the $\sigma_z^k$ operator appears in $\hat P_j$, we do not need to apply any Hadamard operator and, thus, the $2(n+1)$ is the estimated maximum number of Hadamard-like operators needed.
} 
\label{fig_app:exponential_operator}
\end{figure} 
%%%%%%%%%%%%%%%%%%%%%%%%%%%%%%%%%%%%%%%%%%%%%%%%%%%%%%%%%%%%%%%%%%%%%%%%%%%%%%%%%%%%%%%%

Let us consider the implementation of the exponential operator $\exp \{ i \lambda \hat p  \otimes \hat{P}_j \}$ where the Pauli string $\hat{P}_j$ is a product of Pauli operators: $ \hat{P}_j = \prod_i \sigma_i^j$.

Having in mind the implementation in a quantum computer, we can substitute the detector $\hat{p}$ operator with a Pauli operator $\sigma_z^D$.
The exponential operator reads $\exp \{ i \lambda \sigma_z^D  \prod_i \sigma_i^j \}$; therefore, the presence of the detector simply increases the dimension of the Pauli string of one qubit.

To implement the exponential operator, we first rotate the Pauli operator to have a product of $\sigma_z$ operators.
This can be done by applying Hadamard operators $H_{had, x}$ for the $\sigma_x$ and the corresponding $H_{had, y}$ for the $\sigma_y$.
That is $   \left( \prod H_{had, l} \right)\exp \{ i \lambda \sigma_z^D  \hat{P}_j \} \left( \prod H_{had, l} \right)= \exp \{ i \lambda \sigma_z^D \prod_p \sigma_z^p \}$.

The diagonalized exponential operator generates a phase factor $\exp \{\pm i \lambda \}$ depending on the logical string to whom it is applied.
More specifically, taking $\sigma_z \ket{0} = - \ket{0}$ and $\sigma_z \ket{1} = \ket{1}$, the transformation is $\ket{x_1 x_2 ... x_n} \rightarrow \exp \{ i \lambda \} \ket{x_1 x_2 ... x_n}$ if the number of $x_i=0$ is even and  $\ket{x_1 x_2 ... x_n} \rightarrow \exp \{- i \lambda \} \ket{x_1 x_2 ... x_n}$ if the number of $x_i=0$ is odd.

To count the parity of $x_i=0$ in a string, we add an {\it ancilla} qubit in position $n+2$ (the $n+1$ qubit is the detector) and apply a sequence of $n$ ${\rm C}_i {\rm NOT}_{n+2}$ logical gate with $i=1,2, ..., n+1$.
The ${\rm C}_i {\rm NOT}_{n+2}$ is controlled on the $i$-th logical qubit and acts always in the {\it ancilla} qubit.
As a consequence, if we have an even number of $x_i=0$, the {\it ancilla} qubit will end up in the $\ket{0}$ state, otherwise, i.e., for an odd number of $x_i=0$, in the $\ket{1}$ state.
Then we apply a phase gate on the ancilla qubit $\exp \{ i \lambda \sigma_z^a  \}$ and reverse the ${\rm C}_i {\rm NOT}_{n+2}$ operations to reset the {\it ancilla} qubit to its original state $\ket{0}$.
The full circuit is shown in Fig.~\ref{fig_app:exponential_operator}.

To implement this operator we need $2(n+1)$ Hadamard gates, $2(n+1)$ CNOT gates and a phase gate for a total of $4(n+1)+1\approx 4n$ elementary gates for $n \gg 1$.
Notice that this is the worst case where we have to apply $2 n$ Hadamard gates.
In many cases, the Pauli string operators are the product of a limited number of Pauli operators so the number of Hadamard gates can be significantly reduced.
For example, in physical problems where we need to find the energy of the ground state of a Hamiltonian, the Pauli strings are usually the product of only four Pauli operators \cite{McClean2018, Wecker2015}.

For the DM approach, we have an analogous situation.
In quantum computers, the measurement is performed in a fixed basis (usually the one in which $\sigma_z$ is diagonal). 
Since we cannot directly measure an alternative Pauli observable ($\sigma_x$ or $\sigma_y$), we have to rotate the qubit state and perform the measurement in the predetermined basis.
As above, this is done by applying an Hadarmard $H_{had,l}$ operator.
Therefore, to measure the average of a Pauli string, we need a maximum of $n$ $H_{had,l}$ logical gates.

%%%%%%%%%%%%%%%%%%%%%%%%%%%%%%%%%%%%%%%%%%%%%%%%%%%%%%%%%%%%
%\bibliographystyle{apsrev4-1}
%\bibliography{QNDM-bibliography}
%%%%%%%%%%%%%%%%%%%%%%%%%%%%%%%%%%%%%%%%%%%%%%%%%%%%%%%

%%%%%%%%%%%%%%%%%%%%%%%%%%%%%%%%%%%%%%%%%%%%%%%%%%%%%%%%%%%%
% Bibliography
%%%%%%%%%%%%%%%%%%%%%%%%%%%%%%%%%%%%%%%%%%%%%%%%%%%%%%%
%

%%%%%%%%%%%%%%%%%%%%%%%%%%%%%%%%%%%%%%%%%%%%%%%%%%%%%%%

\end{document}